\documentclass[aip,amssymb,amsfonts,amsmath,floatfix,reprint]{revtex4-2}

\usepackage[utf8]{inputenc}
\usepackage{dcolumn}
\usepackage{graphicx}
\usepackage{bm}
\usepackage[bookmarks=false, pdfstartview={FitH}, colorlinks = true, citecolor = blue, linkcolor = blue, urlcolor = blue,hyperfootnotes=true]{hyperref}

\DeclareFontFamily{U}{euc}{}
\DeclareFontShape{U}{euc}{m}{n}{<-6>eurm5<6-8>eurm7<8->eurm10}{}%
\DeclareSymbolFont{AMSc}{U}{euc}{m}{n} 
\DeclareMathSymbol{\umu}{\mathord}{AMSc}{"16}

\def\MEAM{\small Department of Mechanical Engineering \& Applied Mechanics, University of Pennsylvania, Philadelphia, PA 19104, USA}
\def\Phys{\small Department of Physics \& Astronomy, University of Pennsylvania, Philadelphia, PA 19104, USA}
\def\Geo{\small Department of Earth \& Environmental Science, University of Pennsylvania, Philadelphia, PA 19104, USA}

\begin{document}


\title{Flow and aerosol dispersion from wind musical instruments}

\author{Quentin Brosseau}
\thanks{Q.B. and R.R. contributed equally to this work.}
\affiliation{\MEAM}
\author{Ranjiangshang Ran}
\thanks{Q.B. and R.R. contributed equally to this work.}
\affiliation{\MEAM}
\author{Ian Graham}
\affiliation{\Phys}
\author{Douglas J. Jerolmack}
\affiliation{\MEAM}
\affiliation{\Geo}
\author{Paulo E. Arratia}
\thanks{E-mail address: \url{parratia@seas.upenn.edu}}
\affiliation{\MEAM}

\date{\today}

\begin{abstract}
In the midst of the COVID-19 pandemic, many live musical activities had to be postponed and even canceled to protect musicians and audience. Orchestral ensembles face a particular challenge of contamination because they are personnel heavy and instrumentally diverse. A chief concern is whether wind instruments are vectors of contamination through aerosol dispersion. This study, made possible by the participation of members of the Philadelphia Orchestra, brings insight on the modes of production and early life of aerosols of human origin emitted by wind instruments. We find that these instruments produce aerosol levels that are comparable to normal speech in quantity and size distribution. However, the exit jet flow speeds are much lower than violent expiratory events (coughing, sneezing). For most wind instruments, the flow decays to background indoor-air levels at approximately 2 meters away from the instrument's opening. Long range aerosol dispersion is thus via ambient air currents.
\end{abstract}

\maketitle

\section{Introduction}



The COVID-19 pandemic has disrupted many aspects of our daily lives. Live entertainment, for example, has experienced severe restrictions resulting (in some cases) in the temporary disbanding of large bands, orchestras, and choirs. Numerous live performances and festivals had to be postponed and even canceled to avoid the spreading of the disease \cite{Argonne_Study, Newman2020, Garcia2021, Wiki_music}. To date, orchestras are still experimenting with new ways of maintaining a high level of artistic production while keeping their audience and personnel safe. Mainly, they have resorted to performing with remote or limited audiences, adapting their repertoire to promote pieces featuring strings, and have made significant changes in the number of musicians and their position in the auditorium \cite{Tommasini2020, Hedworth2021}.

\begin{figure*}[t]
\centering
\includegraphics[width=4.5in]{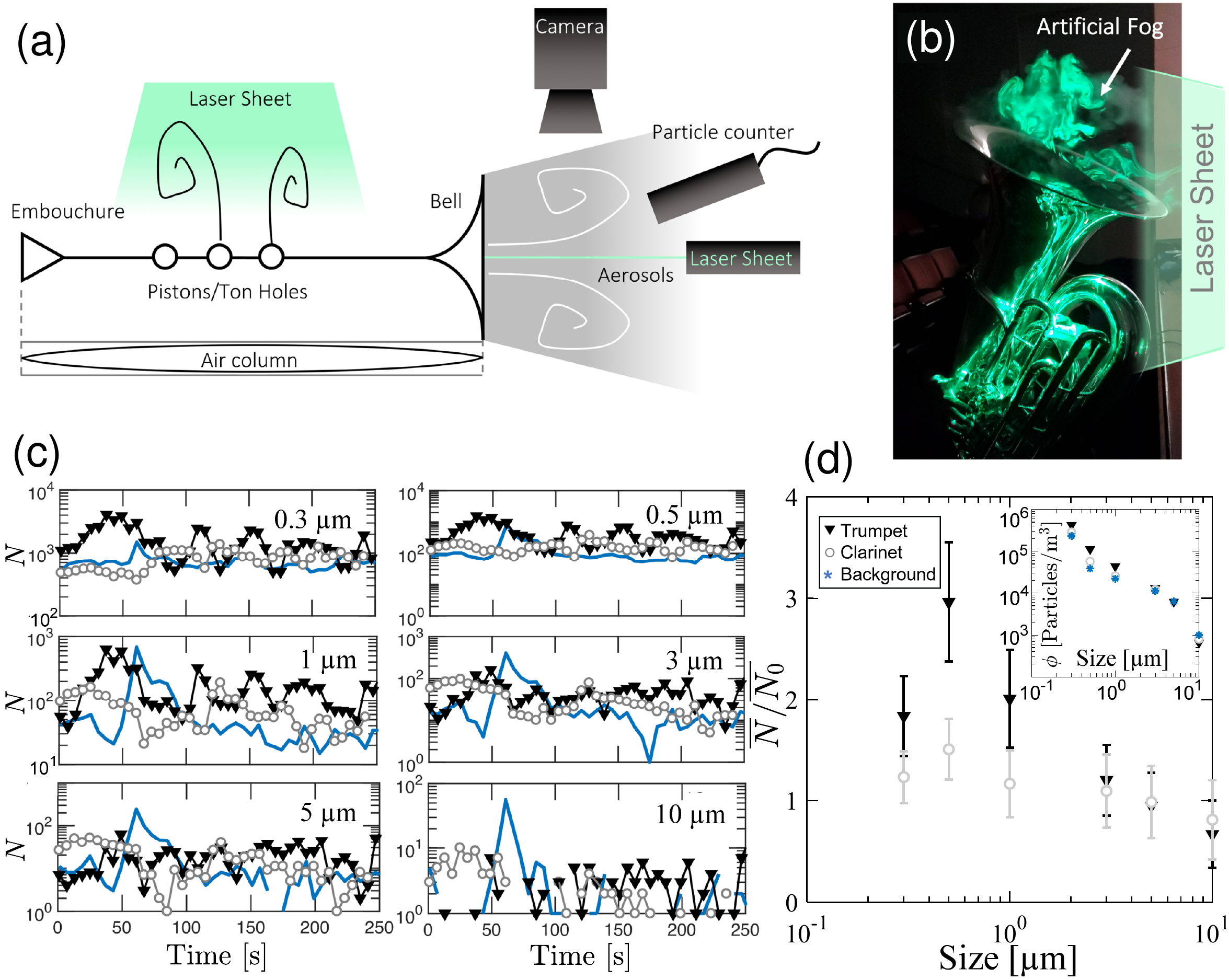}
\vspace{-2mm}
\caption{{\bf{Experimental setup, flow structure, \& aerosol concentration data.}} ({\bf{a}}) Sketch of experimental setup. Two main experiments are performed: flow visualization and particle counting. These experiments are performed separately to avoid contamination. Particle/aerosol production is measured at the bell of wind instruments. Flow structures are visualized with artificial fog at different openings (bell for brass instruments, and tone holes and labium for woodwinds) using a laser sheet and a high-speed camera. ({\bf{b}}) Flow structures produced by the tuba using green laser. ({\bf{c}}) Sample records of the (time-dependent) particle/aerosol emission at the bell for the trumpet (inverted triangles) and the clarinet (hollow circle) compared with background (solid line) in discrete size channel 0.3$~\umu$m, 0.5$~\umu$m, 1$~\umu$m, 3$~\umu$m, 5$~\umu$m, 10$~\umu$m. Acquisition time is 250~s and sampling rate is 6~s. ({\bf{d}}) Time-averaged number of aerosols $\overline{N}$ normalized by the time-averaged background signal $\overline{N_0}$, and the corresponding aerosol concentration (inset), as a function of aerosol size. The error bar corresponds to the standard error.}
\label{fig:setup}
\end{figure*}

Among professionals, musicians playing string instruments and percussion can maintain an activity fully compliant with the recommendations of the World Health Organization (WHO) regarding social distancing and the use of personal protections equipment (face masks, screens, etc.) \cite{WHO_web, feng_lancet_2020,Sha_PoF_2021}. The case of wind instruments, however, poses a unique challenge since they are actuated by the human respiratory system. The constant air stream forced in and out of the body, required to maintain basic tone quality while playing a wind instrument, results in the creation and expulsion of small aerosols from the body. In addition, mouth and tongue movement required for the interpretative aspect of the sound imposes mechanical constraints that further contribute to the production of aerosol particles \cite{Zhou2021}. Combined, these two process can lead to an enhancement of aerosol flux exiting the instrument.


Even in the absence of audiences, the use of wind musical instruments may still pose risks to other musicians as they are likely to release airborne particles, especially given the size of modern orchestra and standard choirs ($\sim100$ persons). Airborne particle transport is usually defined as the spreading of pathogens by aerosols of a size smaller than $5~\umu$m, which allows them to travel over a long distance carried by air flow \cite{Wilson2020, Bhardwaj_POF2021,Bush_PNAS2021, Abhishek_PoF202, Clarke_PoF2020}. This transport process is regarded as the main contamination pathway of the SARS-CoV-2 virus \cite{Lednicky2020, Sommerstein2020, mittal_JFM_2020,Sznitman_Science_2020, Bourouiba_ARFM_2021}. Studies have established that the vast majority of the aerosol generated during normal speech is under the size of $1~\umu$m \cite{Papinieni1997, Asadi2019, Morawska2009}. But larger size ($> 100~\umu$m) particles, which travel ballistically at speeds of 5 to 10 m/s \cite{Xie2009, Bourouiba2014,Han2021} during coughing and sneezing, can still be a factor in contamination especially over shorter distances \cite{Jones2015,Sznitman_Science_2020, mittal_JFM_2020,bourouiba_turbulent_2020}. For wind musical instruments, studies have found a particle size distribution similar to normal speech ($<5~\umu$m) \cite{He2021}, indicating the potential for long-range transport. Moreover, recent experiments have shown the the flow  created by some wind instruments can extend to approximately $30~$cm away from the instrument \cite{Becher2021}; ambient flow remains undisturbed beyond the distance of 1.5 m from the instrument \cite{Spahn2021}. Despite these advancements, there is still a dearth of relevant experimental studies regarding the production and transport of aerosols in wind instruments. 

In this study, we investigate the flow and (mass) flux of aerosols produced by wind musical instruments using flow visualization techniques and particle counting instruments. We focus on single instruments and combine flow with particle concentration measurements to gain insights into aerosols' early life in the vicinity of a single musician. The data are used to formulate predictions regarding aerosol spreading. (Details regarding the dependency on articulation such as pitch, note attack and  intensity of playing can be found in separate studies \cite{He2021, Abraham2021}.) We find that the flow produced by wind instruments is, in general, relatively gentle and that the aerosol fluxes produced are similar to those generated by normal speech. Flow visualization reveals that intermittent jets are produced at the instrument bells or tone holes. Using a simple model, we show that the jet velocity decays to the indoor background flow within a few meters.




\section{Experimental Methods}
\vspace{-2mm}
We begin by defining a few terms specific to musical instruments of the wind class. The air column of a wind instrument is the resonating space contained within the bore of the instruments (Fig. \ref{fig:setup}a). It is bounded by the physical walls made of metallic alloys or wood essences; these wall are designed to be impermeable to air under playing conditions \cite{Bucur2018}. The air column starts at the embouchure (the player's mouth) and ends at the bell, which is an opening, usually with a flared geometry, at the instrument's bell or exit. In modern instruments, bore walls are usually pierced with tone holes or pistons that allow changes in the length of the air column.

All experiments are performed with professional musicians from the Philadelphia Orchestra. To minimize the effects of human activities (e.g. breathing) on the background flow and aerosol generation, musicians were isolated in a large amphitheater (along with the operators). For each measurement, musicians were asked to play their instrument continuously and repeat the same cycle of notes for approximately 2 minutes. We measure aerosol concentration and size distribution, as well as flow velocity in the vicinity of a given instrument using particle counting and flow velocimetry; more details can be found below.

Aerosol detection is performed using a particle counter (Lighthouse Handheld 3016), that uses light scattering technology for particle detection and measurement. The counter abides by ISO 21501-4 calibration standards, which tolerates a size resolution of up to 15 \% and error of up to 10\% of the particle target size. The particle counter performs measurement in 6 discreet channels from 0.3 to 10$~\umu$m. The instrument operates at a pumping rate of 2.83 L/min and a refreshment rate of 6 seconds. These values are used to compute the time-averaged aerosol particle concentration, that, number of particles per unit of volume.

The opening of the aerosol counter is set at a distance of 5 - 20 cm from the instrument bell and tone holes. For each experiment, the musicians are asked to play up and down a scale, in the natural range of the instrument, at a comfortable tempo, and at a mezzo forte dynamic range. The musical notes played can be found in Supplementary Fig. 1. The experiments are purposefully performed in a large amphitheater to reproduce performance conditions, and to avoid the accumulation of released aerosols near the area of the musician. Background measurements are performed before any musician takes place in the auditorium to avoid contamination by human activity in the space. A delay of 20 minutes in between each musician is allowed to renew the air volume in the space.

Flow characterization experiments are performed at the bell for brass instruments and in the region above tone holes for woodwinds, with a custom particle image velocimetry (PIV) setup similar to a previous study \cite{Li2021}. An artificial fog is produced by a commercial ultrasonic humidifier (AquaOasis) and directed towards the instrument (Fig. \ref{fig:setup}b). The fog droplet average diameter is approximately 5 $\umu$m, and their average sedimentation time scale is roughly 30 minutes, which is much larger than the experimental time scale ($<2$ minutes). The droplets behave as passive tracers of the flow since their Stokes number is approximately $10^{-4}$ (see the SI for more details). A slice of the fog parallel to the axis of the bell is illuminated by a green laser sheet (517 nm, 50 mW, BioRay). A camera equipped with a lens (AF-D, 80-200 mm, Nikon) records images of the fog at a rate of 60 fps while each musician performed continuous scales up and down in pitch at a comfortable tempo for a duration of 2 minutes. We note that the focus of this study is on a single instrument; ensemble studies will be the subject of future studies. 

\section{Results and Discussion}
\vspace{-2mm}
\subsection{Aerosol Production}
\vspace{-2mm}
Aerosol concentration and size distribution produced by each wind musical instruments are characterized using a particle counter. Figure \ref{fig:setup}c shows sample records of aerosol counts, $N$, as a function of time ($\Delta t=250$~s) for one representative instrument of the brass family (b-flat trumpet) and one of the woodwinds family (b-flat clarinet). Results are shown for 6 discrete particle diameters ($d=0.3~\umu$m, 0.5$~\umu$m, 1$~\umu$m, 3$~\umu$m, 5$~\umu$m, 10$~\umu$m); also shown is a sample record background count for control. For both instruments, the particle count time signals show deviation from background signal for sizes $\leq 1~\umu$m. We find that the aerosol production rate (number of aerosol per unit time) deceases as aerosol size increases by several orders of magnitude; the signals become indistinguishable, albeit lower than the background for sizes $\geq 5~\umu$m. The data indicate that most of the aerosol is produced in the range of $0.3~\umu$m to $1~\umu$m. 

To quantify the above observations, we compute a normalized time-averaged aerosol count $\overline{N}/\overline{N_0}$, where $\overline{N_0}$ is the background count averaged over many runs before any experiment is performed (Fig. \ref{fig:setup}d). We find that $\overline{N}/\overline{N_0}$ is non-monotonic in aerosol size, $d$, with a peak at 0.5$~\umu$m for both instruments. We note that non-monotonic size distribution in aerosol production has also been observed in activities based on mouth vocalization \cite{Johnson2011, Asadi2019, Morawska2009, Xie2009}. We find that the averaged number of aerosols produced by the trumpet (at the peak) is nearly twice as much as the clarinet. The count signal for the larger particle size bin ($d=10~\umu$m) can at times be below background noise ($\overline{N}/\overline{N_0}<1$), indicating that the signal is weak and buried in noise. Aerosol emission data, $\phi$ (particles/m$^3$), also show that the trumpet produces more aerosols than the clarinet (Fig. \ref{fig:setup}d, inset), in agreement with recent results \cite{He2021}. This may be due to instrument construction. The clarinet, similar to other woodwinds, also possesses a large number of tone holes through which part of the flow is expelled \cite{Becher2021}, thus reducing the concentration of aerosol at the bell.  

For both trumpet and clarinet instruments, we find that most of the aerosols emitted are in the smaller, sub-micrometer ($\le 1~\umu$m) range. This phenomenology is similar to normal respiration ($10^{4}$~m${}^{-3}$) and vocalization ($10^{5}$~m${}^{-3}$) \cite{Morawska2009, McCarthy2021}, indicating that our data for these two specific instruments are more consistent with values obtained by the processes involving vocal cords rather than mouth movements (lips of tongue) such as reading or normal breathing. Other studies find flute and oboe emission ranges to be differing up to two order of magnitude more than speech, suggesting that the playing of these instruments (air pressure and mouth shape) can concentrate  the amount of aerosol produced \cite{Firle2021, He2021}. Finally, our measurements indicate that aerosol distribution and concentration values are not affected by the transit inside the instrument bore for particle sizes $\leq 10~\umu$m since the values are similar to normal speech \cite{Asadi2019, Morawska2009}. It seems that moisture deposition inside bores is a process involving larger aerosol sizes.  



\begin{figure}[t!]
\centerline{\includegraphics[width=3.37in]{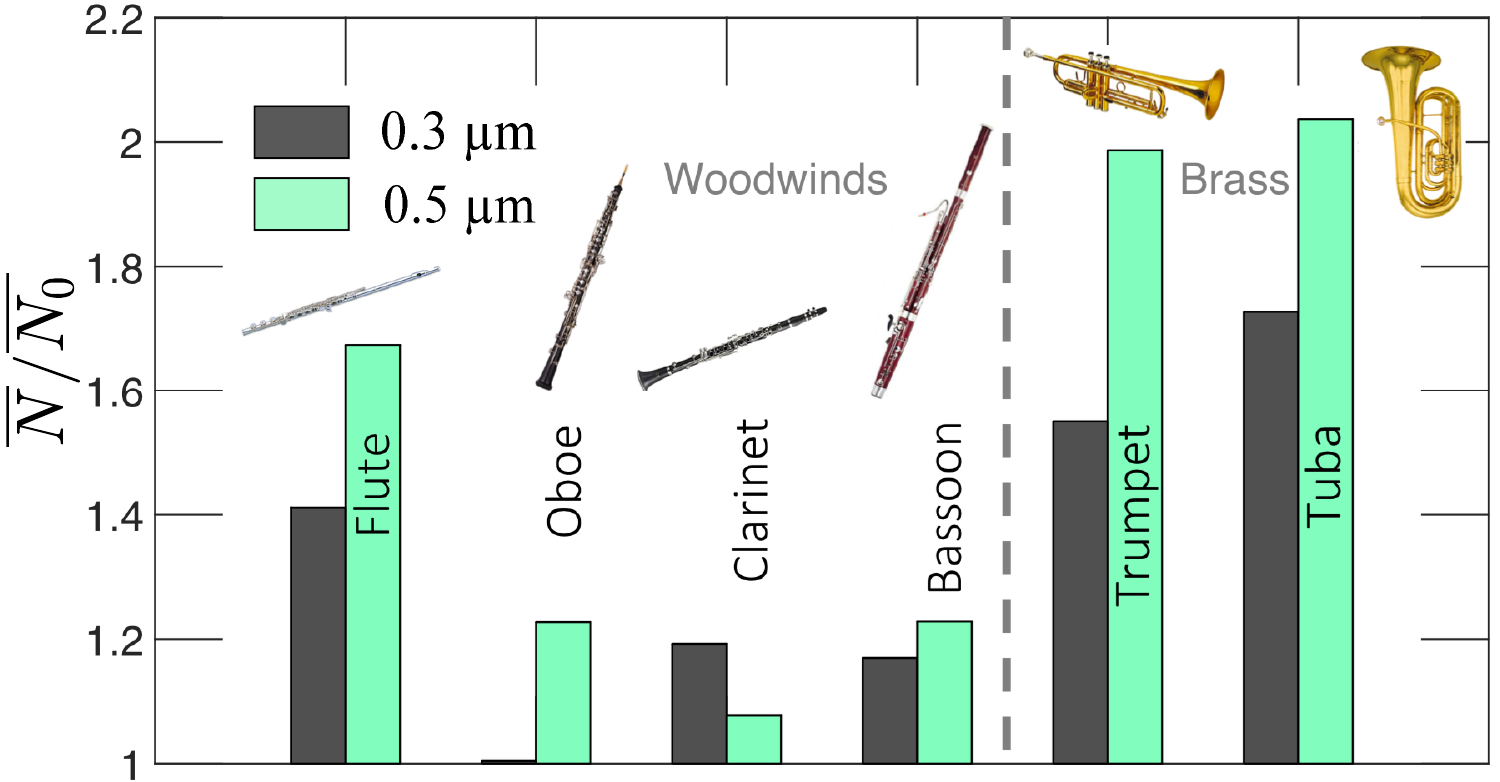}}
\caption{\footnotesize  {\bf{Aerosol production by wind instruments.}} Time-averaged number of aerosols $\overline{N}$ produced by different woodwinds and brass instruments (separated by the dashed line) for aerosols diameters of 0.3 $\umu$m and 0.5 $\umu$m. Data are normalized by the time-averaged background signal $\overline{N_0}$. Brass instruments produce larger amounts of aerosols than woodwinds.
}
\label{fig:aerosols}
\end{figure}



\begin{figure}[b!]
\centerline{\includegraphics[width=3.37in]{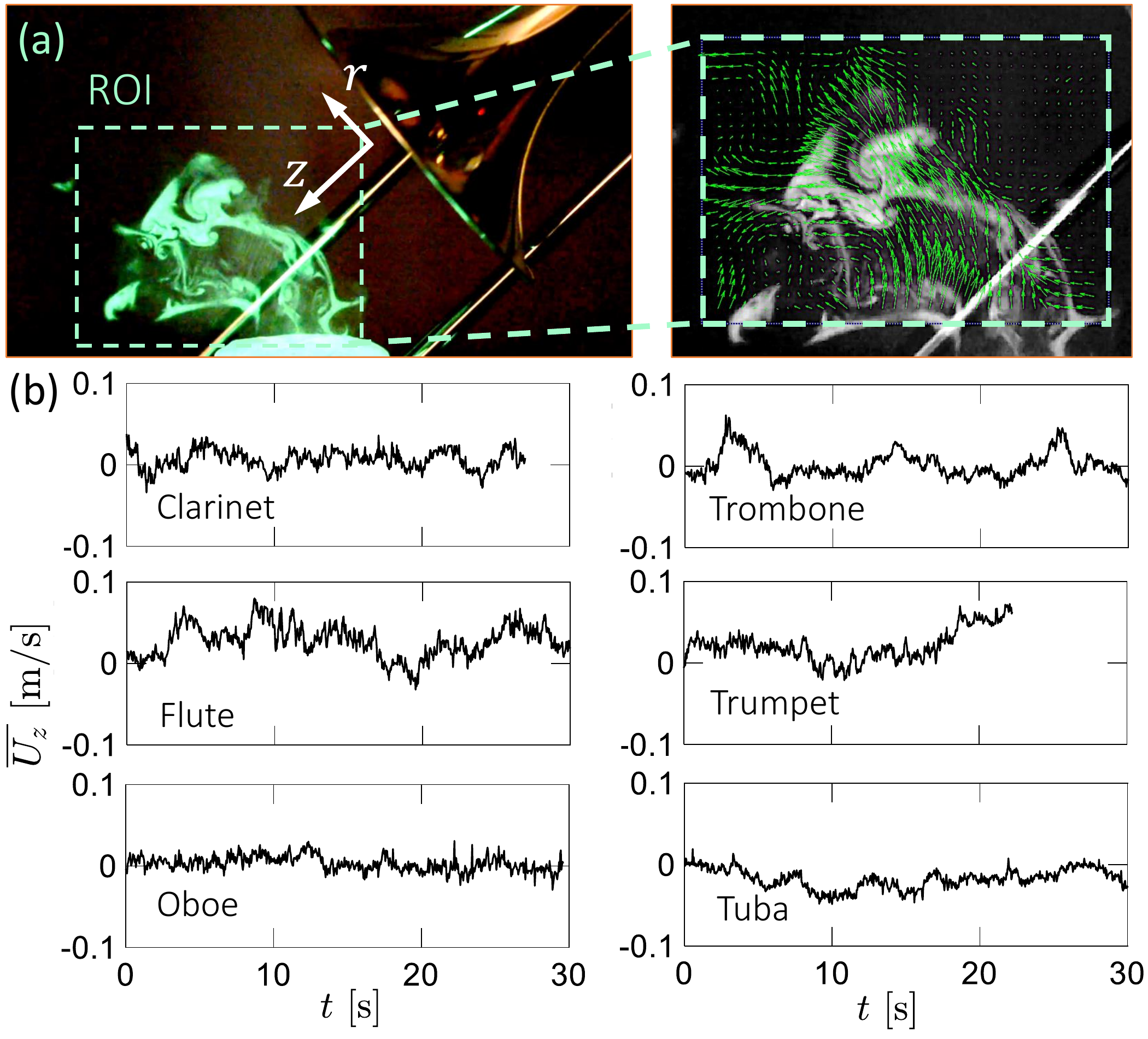}}
\vspace{-2mm}
\caption{\footnotesize {\textbf{Flow characterization \& exit flow velocity}.} ({\textbf{a}}) Laser-induced flow visualisation (left, Multimedia view) and velocity field (right) obtained from PIV for the trombone during play. Also shown are the region of interest (ROI) for the PIV analysis and the coordinate system, where $z$ is the axial direction parallel to the centerline of the instrument's opening, and $r$ is the radial direction normal to the centerline. ({\textbf{b}}) Spatially average axial velocity $\overline{U_z}$ in the ROIs above the woodwind tone holes (clarinet, flute, oboe) and at the bells of brass instruments (trombone, trumpet and tuba) as a function of time. The flow speed of wind instruments is much lower than that of sneezing and coughing (see text). 
}
\label{fig:flow}
\end{figure}

Next, we compare aerosol emission by a broader set of wind instruments. Figure \ref{fig:aerosols} shows the normalized mean aerosol count (or emission), $N/N_0$, for an extended sample of instruments; only the $0.3~\umu$m and $0.5~\umu$m channels are presented here since they are the most abundant. Data are shown for woodwinds (oboe, b-flat clarinet and bassoon) characterized by small tube length, wooden walls and tone holes; and brass (b-flat trumpet and f-tuba), with piston and large tube length. Similar to our previous results, we find that aerosol size distributions for all instruments peak at approximately $d=0.5~\umu$m, except the clarinet, which peaks at $d=0.3~\umu$m. The emission reported here is roughly 1.5 to 3 times that of the background of the room. Surprisingly, for the selected size range (0.3 - $0.5~\umu$m), the amount of aerosol emitted is independent of the instrument's tube length. For example, the tuba, with its large cross section and tube (length $\sim 5.5$~m), has a comparable emission rate to the trumpet tube (length $\sim 1.5$~m) which itself is twice as long as the clarinet (length $\sim 70$~cm). Note that the larger volume of air needed for the playing tuba could explain the larger amount of aerosols. This is further evidence that deposition of droplet on the instrument wall is minimal for this size range. We note that previous studies on aerosol generation from normal speech show additional modes below 300 nanometers \cite{Morawska2009,Johnson2011,Scheuch2020}, which is beyond our current resolution.  

Overall, our results show wind instruments emit aerosols mostly in the sub-micrometer range ($\leq 1~\umu$m) at concentrations comparable to normal respiration and vocalization. Particle counting data show that woodwind instruments are lower emitters relative to brass instruments. However, the flute stands apart; it has particle concentrations similar to brass instruments, albeit lower. Note that the flute's aerosol measurements are performed next to the instrumentalist's mouth above the mouthpiece, since this is the most significant stream exiting the instrument \cite{Benson2006,Fletcher1998}. Thus, the flute data are not affected by possible losses through the tone holes. The question now is: how far do these aerosols travel? To address this issue, we characterize the flow emanating from the musical instruments in the following section.     


\subsection{Flow Characterization}
\vspace{-2mm}

We now characterize the flow at the openings (i.e. bells and tone holes) of wind musical instruments using a particle image velocimetry (PIV) technique with a high-speed camera. In short, fog is used as seeding particles that are visualized using a laser sheet (see Fig. \ref{fig:setup}b and Supplementary Fig. 2). Flow velocities are obtained by the cross-correlation between two successive images taken by the camera in the region of interest (ROI). The size of the ROI is commensurate with the diameter of the instrument's opening. Figure \ref{fig:flow}a shows an example of a flow field extracted at the bell of a trombone. The coordinate system is defined such that the axial direction $z$ is parallel to the centerline of the instrument's opening, and the radial direction $r$ is normal to the centerline (see Fig. \ref{fig:flow}a). In this case, the velocity in the axial direction, $U_z$, characterizes the directional flow exiting the instrument's opening. Each time-dependent flow field is then spatially averaged to obtain the average axial velocity, $\overline{U_z}$, as a function of time for each instrument.

Figure \ref{fig:flow}b shows $\overline{U_z}$ as a function of time for all instruments analyzed here. In general, the values of $\overline{U_z}$ are relatively similar for all instruments. In fact, our measurements do not show a distinct correlation between the pitch evolution (up of down a musical scale) and the value of the mean velocity; see the SI for more details pitch evolution. Importantly, these flow speeds ($<0.1$ m/s) are relatively low compared to the speeds of coughing and sneezing ($\sim5$ - 10 m/s) \cite{Bourouiba2014}. Even for the flute, the highest instantaneous average velocity (0.07~m/s) is an order of magnitude lower than the speeds for coughing and sneezing \cite{Han2021, de_Silva2021}.

\begin{figure}[b!]
\centerline{\includegraphics[width=3.37in]{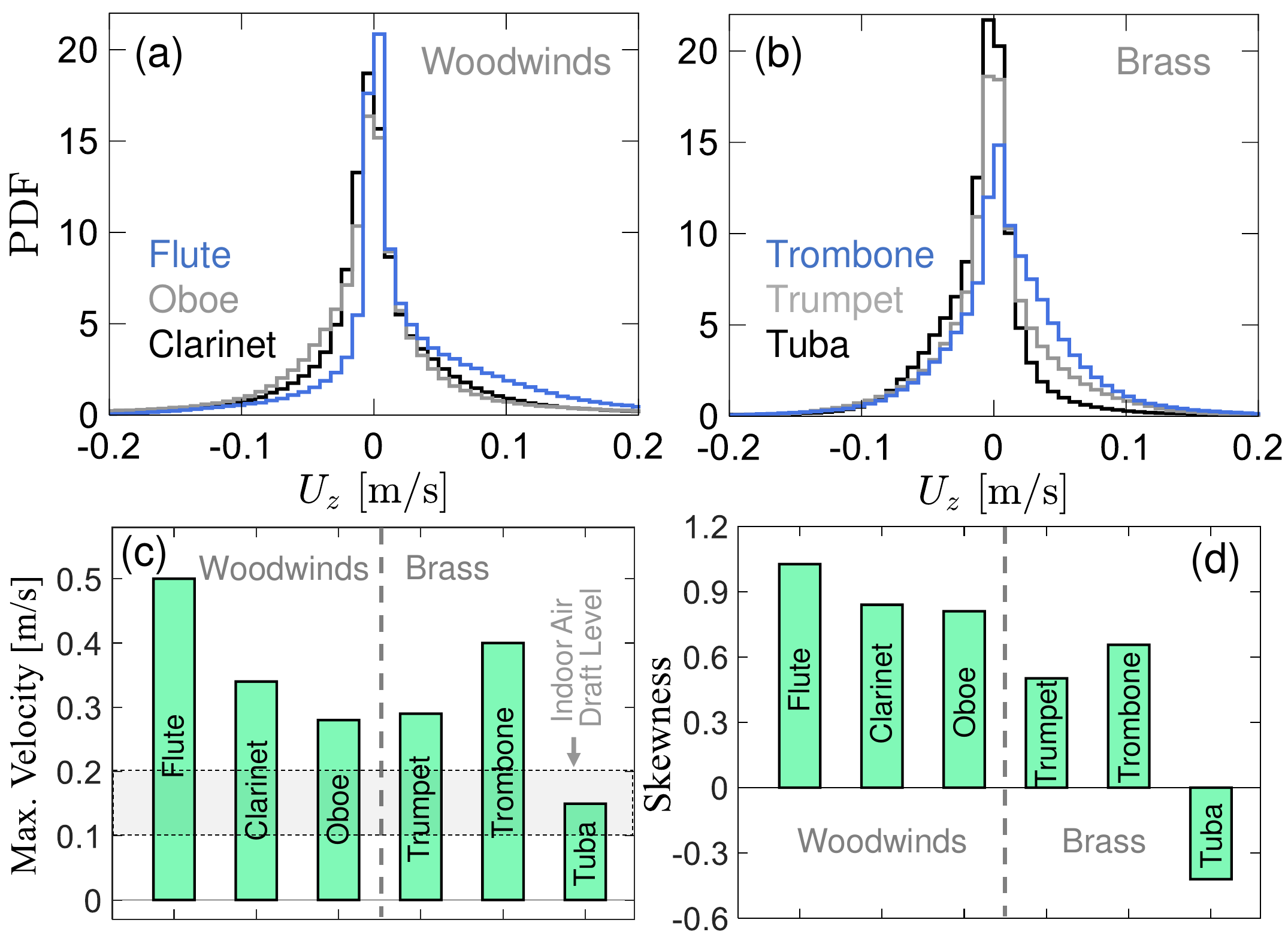}}
\caption{\footnotesize  {\textbf{Flow statistics}} ({\textbf{a, b}}) Probability density functions (PDF) of axial velocity $U_z$ for woodwinds in \textbf{a} and brass in \textbf{b}. ({\textbf{c}}) The maximum axial velocity, defined by the 99-percentile of the velocity distribution, shows the maximum speed at which the exit jet travels. ({\textbf{d}}) The skewness of the PDF of $U_z$, characterizing velocity bias away from the instrument opening. The flute shows the strongest directional flow of all the wind instruments. 
}
\label{fig:stats}
\end{figure}



To gain further insight into the flow generated by the instruments, we compute and analyze the probability distribution of $U_{z}$ obtained from PIV over the duration of the recordings (Fig. \ref{fig:stats}a, \ref{fig:stats}b). For all the instruments, the mean of the velocity distribution is of the order of $10^{-2}$ m/s, indicating that the flow is quite gentle in accordance to recent results \cite{Spahn2021}. The maximum velocity, defined as the upper 99-percentile of the distribution, reaches as large as the order of magnitude of $10^{-1}$ m/s (Fig. \ref{fig:stats}c), which are comparable to the velocity of normal respiration events ($\sim 1$~m/s) \cite{Tang2013}, but small in comparison to coughing and sneezing events \cite{Han2021, de_Silva2021}. 

The skewness ($Sk$) of each distributions has a visible bias towards positive values. This indicates the instruments are likely to produce velocities larger than the mean values in the direction away from the instrument opening. We find that $Sk\sim 1$ for woodwinds and $Sk\sim 0.5$ at the bell of brass instruments, which indicates a stronger directional displacement above tone holes compared to the bell. Curiously, the tuba shows a negative skewness (Fig. \ref{fig:stats}d). Tuba is the only measurement in the upright direction, and it is likely that the negative skewness is due to (fog) droplet sedimentation  in direction opposite to the bell (Fig. \ref{fig:stats}d). Overall, both the skewness and mean velocity data show that rather than a continuous stream, the flow displacement at the opening is better characterized by transient events that dominate the background ambient flow intermittently. The observed intermittency might originate from pressure variations necessary for the inception or "attack" of a pitch \cite{Hoekje2013}, as well as changes in fingering that release or block pistons and tone holes. This results in flow structures that contain puffs and jets \cite{Becher2021, Abraham2021}; see Fig. \ref{fig:setup}b as well as the SI for time lapse images and videos. 

It is worth noting that flutes (together with recorders and organ pipes) stand apart as the vibration of the air columns is set in motion by an air jet instability that originates at the mouth of the musician over the embouchure \cite{Campbell2021, Benson2006}. In that case, our measurements are taken in a direction aligned with the mouth of the musician, instead of above the tone holes. Despite this peculiarity, the flute displays a skewed velocity distribution comparable to other woodwinds. 

\subsection{A simple laminar jet model}
\begin{figure}[b!]
\centerline{\includegraphics[width=3.37in]{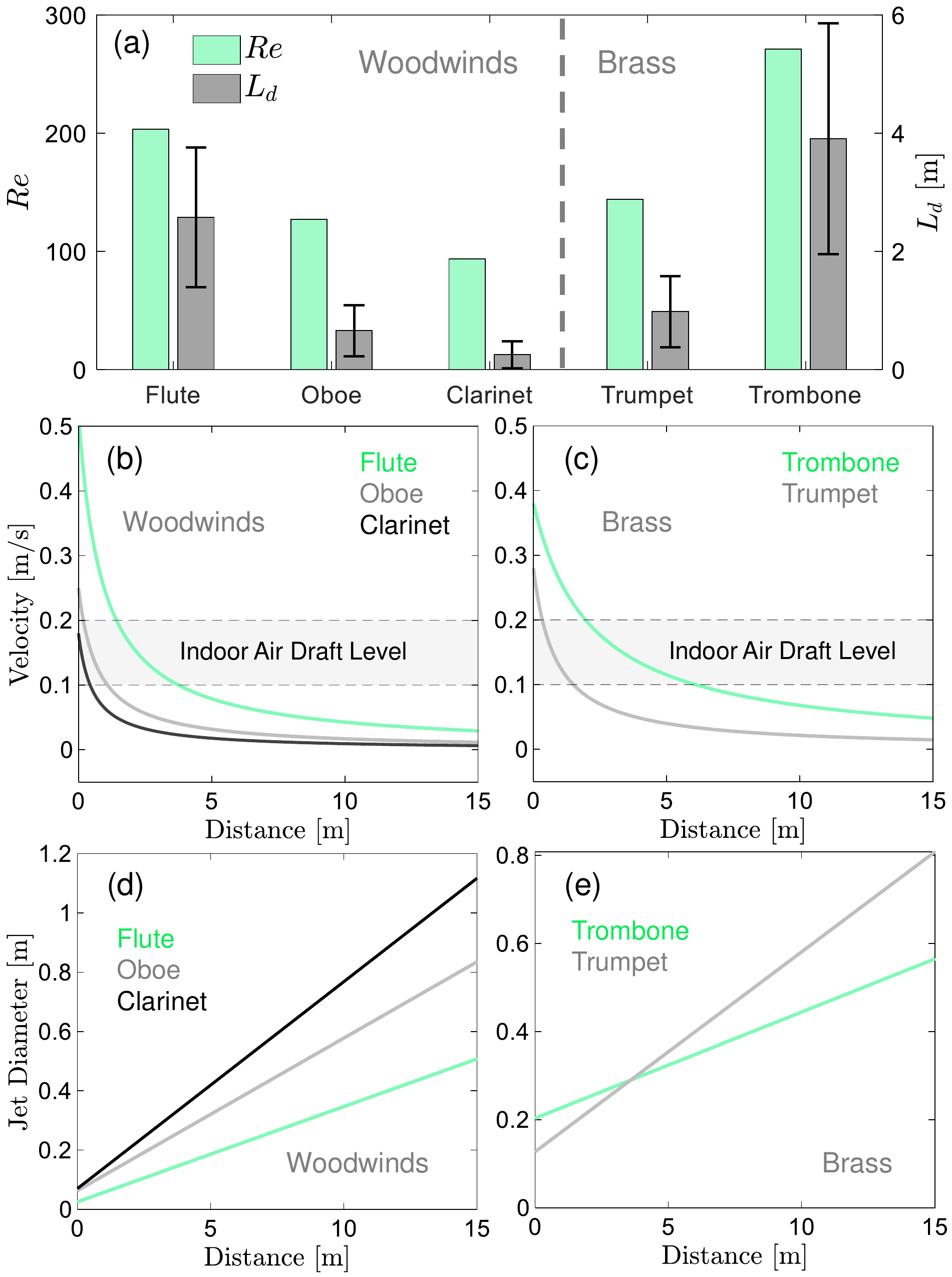}}
\vspace{-2mm}
\caption{\footnotesize {\textbf{Jet velocity decay.}} (\textbf{a}) Maximum Reynolds number $Re$ and jet velocity decay length $L_d$ for different wind instruments. Here, $L_d$ is the distance for the jet centerline velocity to decay to indoor air draft level (0.1 to 0.2~m/s), as shown by the grey area in \textbf{b} and \textbf{c}. (\textbf{b, c}) Axial velocity decay of the jet as a function of distance $z$ from the instrument's exit, for woodwind in \textbf{b} and brass in \textbf{c}. Only flute and trombone can reach beyond 2 meters, based on a conservative estimate using the maximum speed and the lowest air draft level. (\textbf{d, e}) The growth of the jet diameter versus the distance $z$ from the instrument's exit, for woodwind in \textbf{d} and brass in \textbf{e}. The larger the Reynolds number is, the further the jet can travel without much expansion in diameter.
}
\label{fig:jets}
\end{figure}

\vspace{-2mm}
To better understand the release and dispersion of aerosols in the ambient, we analyze the characteristics of the jets emitted by the musical instruments during play (See Supplementary Fig. 2). We begin by calculating the maximum Reynolds number ($Re$) for each instrument. Here, we define $Re=U_mD/\nu$, where $U_m$ is the maximum (or peak) air velocity obtained from PIV, $D$ is the diameter of the instrument's opening (bell or tone hole), and $\nu$ is the kinematic viscosity of air. The maximum air velocity is used to provide a conservative estimation. The (maximum) Reynolds numbers for all instruments are well below 300 (Fig. \ref{fig:jets}a), which indicates that the jet flow is still laminar. Therefore, we model the flow exiting the instrument opening as laminar jets. We further simplify the problem by assuming steady and an axisymmetric jet. Under such conditions, we can make use of an analytical solution known as the ``Schlichting jet'' \cite{Schlichting_1979}. The steady and axisymmetric assumptions are clear simplifications, but they allow us to estimate the decay of jet velocity without resorting to numerical simulations.

The axial velocity of the laminar jet produced by the instruments can be expressed as (see details in the SI):
\begin{equation}\label{Schlichting eqn.}
    U(z,r) = \frac{\nu}{z}\frac{3Re^2/4}{\left(1+3Re^2r^2/32z^2\right)^2},
\end{equation}
where $z$ is the axial direction parallel to the jet flow, $r$ is the radial direction normal to the flow (see Fig. \ref{fig:flow}a), and $Re$ is the Reynolds number of the jet defined in the above manner. The volumetric flow rate of such jet, $Q$, increases with axial distance $z$ due to the entrainment of ambient air:
\begin{equation}
Q = 2\pi\int_0^{\infty}Ur\,dr=8\pi\nu z.
\end{equation}
We note that Equation \eqref{Schlichting eqn.} is valid only for distances far away from the jet origin, i.e., for larger $z$ values. Hence, we select a reference distance, $z_0$, that matches the volumetric flow rate of the instrument such that $Q=8\pi\nu z_0=U_mA$, which leads to:
\begin{equation}
    z_0=\frac{U_mA}{8\pi\nu}=\frac{U_m\pi D^2/4}{8\pi\nu}=\frac{U_mD^2}{32\nu},
\end{equation}
where $A=\pi D^2/4$ is the cross-sectional area of the opening of the instruments. We can then replace the variable $z$ by $(z+z_0)$ and express the centerline velocity of the jet at $r=0$ as:
\begin{equation}\label{eqn. jet}
    U_c = U(z+z_0,0) = \frac{3Re^2\nu}{4(z+z_0)}.
\end{equation}

The above equation describes how the jet axial velocity emanating from the instrument's exit decays as a function of axial distance. It predicts that the jet velocity far away from the exit is inversely proportional to the propagation distance $z$ and proportional to the square of the Reynolds number: $U_c\sim Re^2/z$. Figures \ref{fig:jets}b and \ref{fig:jets}c plot the estimated jet velocity decay versus the (jet) propagation distance for woodwind and brass instruments based both on Equation \eqref{eqn. jet} and PIV measurements. The data show that flute and trombone have significantly higher velocities than the other instruments, as a result of their larger $Re$ values (see Fig. \ref{fig:jets}a) due to either larger opening and/or higher flow speeds. In both cases, the distance needed for the flow jet to decay to the indoor air draft (0.1 - 0.2 m/s) is of the order of a few meters.

But would the aerosol particles follow the flow? To answer this question, we compute the Stokes number ($Stk$), which characterizes the ``attachment'' of a suspended particle to the surrounding flow. For the range of flow speeds measured in this study, we find that the Stokes number are well below unity for all aerosol diameters ($Stk<10^{-2}$, see Supplementary Fig. 4). In this regime, the aerosols act as ideal tracers that closely follow the flow streamlines. As the jet velocity decays to room air draft level, the aerosol particles will be carried away by the ambient air flow.

Thus, we can define a length scale, $L_d$, that represents the farthest distance an aerosol particle can travel with the jet until it is carried away by the ambient air flow. Figure \ref{fig:jets}a shows the values of $L_d$ for all instruments as an average between low (0.1 m/s) and high (0.2 m/s) indoor air draft speeds. Results show that aerosols released by the flute and trombone can propagate over a distance of 2~m to 6~m before reaching the room air draft level. For the other instruments, the decay length ($L_d$) are all shorter than 2~m, the recommended social distancing of the CDC guidelines \cite{CDC_Guideline}. Note that the calculation is performed with the instrument's maximum flow speeds, and thus representing the most conservative estimation. Estimations using the average speeds would lead to much lower distances ($L_d \ll 1$~m).

We are also interested in the dispersion of aerosol particles in the direction normal to the flow. To this end, we estimate the expansion of the jet versus the propagation distance, and find that the jet diameter is proportional to the axial distance $z$, and inversely proportional to the Reynolds number: $D\sim z/Re$  (see the SI). Figures \ref{fig:jets}d and \ref{fig:jets}e show the growth of jet diameter for different wind instruments. We find that the jets generated by instruments with larger $Re$, flute and trombone, can propagate a longer distance without much expansion in diameter. On the other hand, the jets created by instruments with smaller $Re$, clarinet and oboe, are more likely to disperse aerosols at short ranges due to their faster expanding diameters, despite unable to travel over long distances.

\section{Conclusion}
\vspace{-2mm}
We present a characterization and analysis of aerosol emission by wind instruments, in collaboration with musicians from the Philadelphia Orchestra. The combination of flow and particle concentration measurements allow us to draw new insights on the early life of these aerosols, from production to their release in the ambient. Overall, we find that aerosols emitted by wind instruments share similar concentration and size distributions to normal speech and respiration events, albeit higher in some cases; that is, the instrument is the main source of aerosol. We show that for particle sizes $ <10~\umu$m, aerosols travel down the instrument bore with little trapping and are either ejected through tone holes or the instrument bell. Flow measurements (using particle image velocimetry) show that exit jet speeds are much lower ($\sim 0.1$~m/s) than coughing and sneezing ($\sim 10$~m/s) events. An analytical solution for an axisymmetric laminar jet along with velocimetry data are used to estimate each instrument's exit jet speed decay. We find that, for most instruments, the maximum decay length is less than 2 meters away from the instrument's opening. Although simplified, the jet model predicts plume length that are consistent with  studies in controled environment \cite{Becher2021} and CFD modelling \cite{Stockmann2021, Abraham2021} which also predict maximum jet development over a distances shorter than 1 m. Aerosols in the size range explored here ($\leq 10~\umu$m) follow the (jet) flow closely until this decay length ($\approx 2$~m) is reached. Beyond that, we expect aerosols to be dispersed by the ambient flow. 


\section*{Supplementary Information}
See the Supplementary Information (SI) for the videos recording of the instruments playing used for the PIV analysis. The derivation for the laminar jet model and the aerosol particles Stokes number and settling times are also provided in the SI.  

\section*{Acknowledgements}
The authors thank the members of the Philadelphia Orchestra, instrumentalists, and staff for their willingness to participation in our study. We also thank P.J. Brennan and Tanya Derksen for their invaluable guidance. We also thank Nakul Deshpande and Andrew Gunn for insightful discussions and assistance with experiments. R.R. and P.E.A. are supported by NSF-DMR-1709763, and I.G. acknowledges the support from NSF-DMR-1720530 (Penn MRSEC).   

\section*{Author Contributions}
\vspace{-2mm}
D.J.J. and P.E.A. designed research, Q.B. and I.G. performed aerosol characterization, Q.B. and R.R. performed PIV measurements, R.R. and P.E.A. developed jet analytical models, Q.B., R.R., I.G., D.J.J. and P.E.A. wrote the manuscript. Q.B. and R.R contributed equally to this work.

\section*{References}
\vspace{-2mm}
\bibliography{ms.bib}

\end{document}